%% file: main.tex
\def\year{2021}\relax
\documentclass[letterpaper]{article} 
\usepackage{aaai21}  
\usepackage{times}  
\usepackage{helvet} 
\usepackage{courier}  
\usepackage{comment}
\usepackage[hyphens]{url}  
\usepackage{multirow}
\usepackage{times,flushend}
\usepackage{subfigure}
\usepackage[namelimits]{amsmath} 
\usepackage{amssymb}            
\usepackage{amsfonts}  
\usepackage{mathrsfs}  
\usepackage{color}
\usepackage{graphicx} 
\usepackage{natbib}  
\usepackage{caption} 

\usepackage[switch]{lineno}
\title{Dynamic Memory based Attention Network for Sequential Recommendation}
\author {
        Qiaoyu Tan \textsuperscript{\rm 1},
        Jianwei Zhang \textsuperscript{\rm 2},
        Ninghao Liu \textsuperscript{\rm 1}, 
        Xiao Huang \textsuperscript{\rm 3}\\
        Hongxia Yang \textsuperscript{\rm 2},
        Jingren Zhou \textsuperscript{\rm 2},
        Xia Hu \textsuperscript{\rm 1} \\
}
\affiliations {
    \textsuperscript{\rm 1} Texas A\&M University,
    \textsuperscript{\rm 2} Alibaba Group,
    \textsuperscript{\rm 3} The Hong Kong Polytechnic University\\
    \{qytan,nhliu43,xiahu\}@tamu.edu, xiaohuang@comp.polyu.edu.hk\\ \{zhangjianwei.zjw, yang.yhx, jingren.zhou\}@alibaba-inc.com
}

\begin{document}
\maketitle

\begin{abstract}
Sequential recommendation has become increasingly essential in various online services. It aims to model the dynamic preferences of users from their historical interactions and predict their next items. The accumulated user behavior records on real systems could be very long. This rich data brings opportunities to track actual interests of users. 
Prior efforts mainly focus on making recommendations based on relatively recent behaviors. 
However, the overall sequential data may not be effectively utilized, as early interactions might affect users' current choices. Also, it has become intolerable to scan the entire behavior sequence when performing inference for each user, since real-world system requires short response time. To bridge the gap, we propose a novel long sequential recommendation model, called Dynamic Memory-based Attention Network (DMAN). 
It segments the overall long behavior sequence into a series of sub-sequences, then trains the model and maintains a set of memory blocks to preserve long-term interests of users.
To improve memory fidelity, DMAN dynamically abstracts each user's long-term interest into its own memory blocks by minimizing an auxiliary reconstruction loss. 
Based on the dynamic memory, the user's short-term and long-term interests can be explicitly extracted and combined for efficient joint recommendation.
Empirical results over four benchmark datasets demonstrate the superiority of our model in capturing long-term dependency over various state-of-the-art sequential models.

\end{abstract}

\section{Introduction}\label{sec:intro}
\input{1.introduction.tex}
\section{The Proposed DMAN Model}\label{sec:prelim}
\input{3.model.tex}

\section{Experiments and Analysis}\label{sec:experiment}
\input{4.experiment.tex}

\section{Related Work}\label{sec:relatework}
\input{2.relatework.tex}
\section{Conclusions}\label{sec:experiment}
\input{5.conclusion.tex}

\bibliographystyle{ACM-Reference-Format}
\bibliography{LSRec}

\end{document}

%% file: 1.introduction.tex
Recommender systems have become an important tool in various online systems such as E-commerce, social media, and advertising systems to provide personalized services~\cite{hidasi2015session,ying2018graph}. One core stage of live industrial systems is candidate selection and ranking~\cite{covington2016deep}, which is responsible for retrieving a few hundred relevant items from a million or even billion scale corpus. Previously, researchers resort to collaborative filtering approaches~\cite{sarwar2001item} to solve it by assuming that like-minded users tend to exhibit similar preferences on items. Typical examples including models based on matrix factorization~\cite{sarwar2001item}, factorization machines~\cite{rendle2010factorization}, and graph neural networks~\cite{ying2018graph,wang2019neural,tan2020learning}. 
However, these methods ignore the temporal dynamics of user behaviors. 

To capture sequential dynamics for user modeling, various sequential recommendation methods have been proposed to make recommendations based on user's past behaviors~\cite{hidasi2015session,tang2018personalized,tan2021wsdm}. They aim to predict the next item(s) that a user is likely to interact with, given her/his historical interactions. Recently, a myriad of attempts that build upon sequential neural networks, such as recurrent neural network (RNNs), convolutional neural networks (CNNs), and self-attention networks, have achieved promising results in various recommendation scenarios~\cite{hidasi2018recurrent,yuan2019simple,yan2019cosrec,sun2019bert4rec,zhang2018next}. The basic paradigm is to encode a user's historical interactions into a vector using various sequential modules based on the behavior sequence, which is obtained by sorting her/his past behaviors in chronological order.

However, as many E-commerce and social media systems keep accumulating users' records, the behavior sequences have become extraordinarily long. For example, more than twenty thousand customers on the Alibaba e-commerce platform have interacted with over one thousand items from April to September 2018~\cite{ren2019lifelong}. Despite the fertile information contained in these long behavior sequences, existing sequential recommendation algorithms would achieve sub-optimal performance in modeling long behavior sequences. The reason is that standard sequential architectures (e.g., RNNs, CNNs, attention networks) are insufficient to capture long-term dependencies confirmed in sequence learning~\cite{graves2014neural,yu2015multi,dai2019transformer}. Directly applying them to model long behavior sequences would result in significant performance degeneration. Thus, in this paper, we target at exploring sequential recommendation with extraordinary long user behavior sequences.
%

There are three major challenges in learning from long behavior sequences. 1) Given that the response time in real-world systems is limited, it has become expensive to scan over the entire behavior sequence at each prediction time~\cite{zhu2018learning}. Existing sequential recommender systems often require read the whole behavior sequence. A few recent long sequential recommendation methods explore splitting the whole input behavior sequence into short-term and long-term behavior sequences and then explicitly extracting a user's temporal and long-term preferences~\cite{ying2018sequential,lv2019sdm}. Despite their simplicity, they still suffer from high computation complexity since they need to scan over the whole behavior sequence during inference. 2) It is crucial to model the whole behavior sequence for a more accurate recommendation. A few attempts have been made to focus only on short-term actions~\cite{li2020time,hidasi2018recurrent} and abandon previous user behaviors. Nevertheless, studies~\cite{ren2019lifelong,belletti2019quantifying} have demonstrated that user preferences may be influenced by her/his early interactions beyond the short-term behavior sequence. 3) It is hard to explicitly control the contributions of long-term or short-term interests for user modeling. Some studies resort to memory neural network~\cite{graves2014neural} to implicitly preserve the long-term intentions for efficient sequential modeling~\cite{chen2018sequential,ren2019lifelong}. But they may suffer from long-term knowledge forgetting~\cite{sodhani2018training}, due to that the memory is optimized by predicting the next-item. Therefore, an advanced sequential model is needed to explicitly model both long-term and short-term preferences, as well as supporting efficient inference.

To address the limitations above, we propose a novel dynamic memory-based self-attention network, dubbed DMAN, to model long behavior sequence data. It offers standard self-attention networks to capture long-term dependencies for user modeling effectively. To improve model efficiency, DMAN truncates the whole user behavior sequence into several successive sub-sequences and optimizes the model sequence by sequence. Specifically, a recurrent attention network is derived to utilize the correlation between adjacent sequences for short-term interest modeling. Meanwhile, another attention network is introduced to measure dependencies beyond consecutive sequences for long-term interest modeling based on a dynamic memory, which preserves user behaviors before the adjacent sequences. Finally, the two aspect interests are adaptively integrated via a neural gating network for the joint recommendation. To enhance the memory fidelity, we further develop a dynamic memory network to effectively update the memory blocks sequence by sequence using an auxiliary reconstruction loss.  
To summarize, the main contributions of this paper are as follows:
\begin{quote}
\begin{itemize}
\item We propose a dynamic memory-based attention network DMAN for modeling long behavior sequences, 
which conducts an explicit and adaptive user modeling and supports efficient inference. 

\item We derive a dynamic memory network to dynamically abstract a user's long-term interests into an external memory sequence by sequence.

\item Extensive experiments on several challenging benchmarks demonstrate our method's effectiveness in modeling long user behavior data.
\end{itemize}
\end{quote}

\begin{figure*}[h]
  \centering
  \includegraphics[width=17.6 cm,height=6.6 cm]{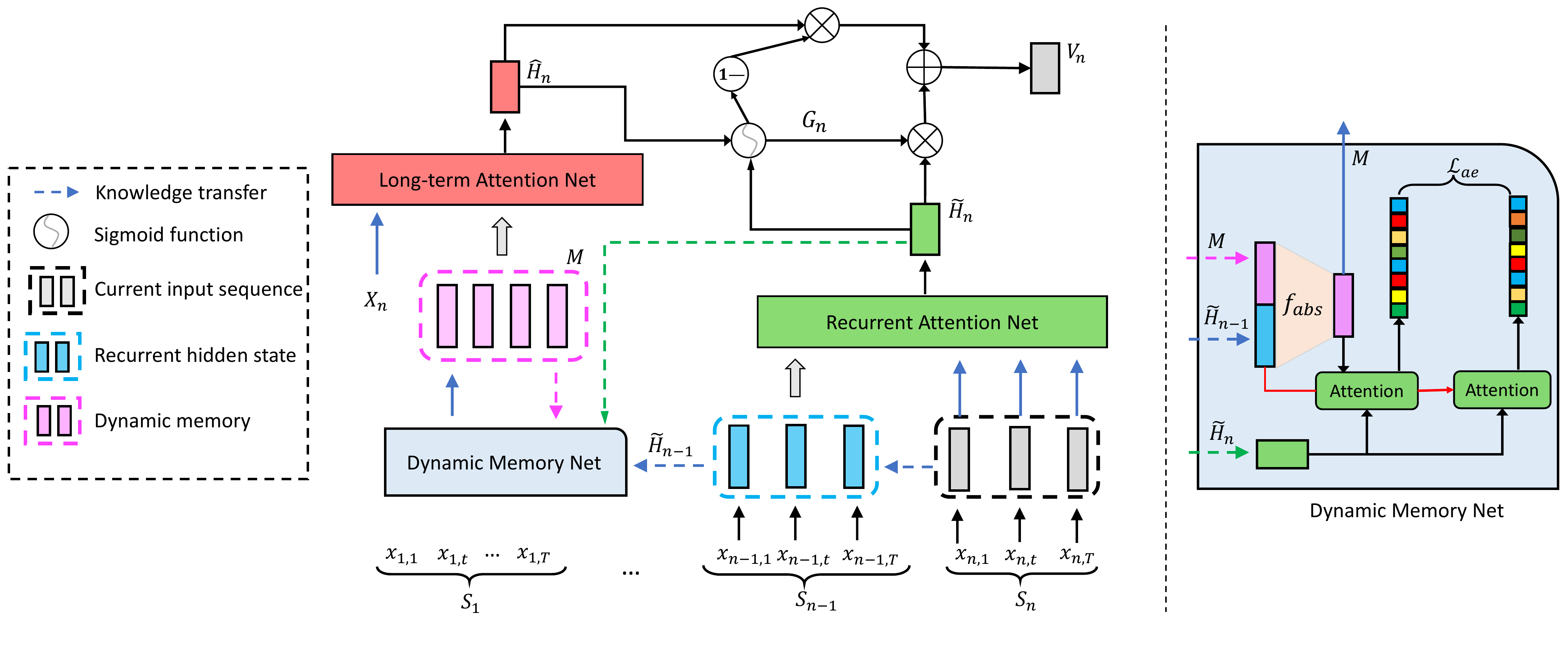}
  \caption{Illustration of DMAN for one layer. It takes a series of sequences as input and trains the model sequence by sequence. When processing the $n$-th sequence $\mathcal{S}_n$, the recurrent attention network is applied to extract short-term user interest by using the previous hidden state $\widetilde{\mathbf{H}}_{n-1}$ as context. Meanwhile, the long-term attention network is utilized to extract long-term interest based on the memory blocks $\mathbf{M}$. Next, the short-term and long-term interests are combined via a neural gating network for joint user modeling. Finally, the dynamic memory network updates the memory blocks via fusing the information in $\widetilde{\mathbf{H}}_{n-1}$, and the model continues to process the next sequence. The overall model is optimized by maximizing the likelihood of observed sequence, while the dynamic memory network is trained based on a local reconstruction loss $\mathcal{L}_{ae}$.
  }
  \label{figure1}
\end{figure*}

%% file: 3.model.tex
In this section, we first introduce the problem formulation and then discuss the proposed framework in detail. 

\subsection{Notations and Problem Formulation}
Assume
$\mathcal{U}$ and $\mathcal{V}$ denote the sets of users and items, respectively. 
$\mathcal{S}=\{\mathbf{x}_1,\mathbf{x}_2,\ldots,\mathbf{x}_{|\mathcal{S}|}\}$ represents the behavior sequence in chronological order of a user. $\mathbf{x}_t\in\mathcal{V}$ records the $t$-th item interacted by the user. 
Given an observed behavior sequence $\{\mathbf{x}_1,\mathbf{x}_2,\ldots,\mathbf{x}_{t}\}$, 
the sequential recommendation task is to predict the next items that the user might be interacted with. Notations are summarized in Table~\ref{table:notation}.

In our setting, due to the accumulated behavior sequence $\mathcal{S}$ is very long, we truncate it into a series of successive sub-sequences with fixed window size $T$, i.e., $\mathcal{S}=\{\mathcal{S}_n\}_{n=1}^N$, for the model to process efficiently. 
$\mathcal{S}_n=\{\mathbf{x}_{n,1},\mathbf{x}_{n,2},\ldots,\mathbf{x}_{n,T}\}$ denotes the $n$-th sequence. Traditional sequential recommendation methods mainly rely on a few recent behaviors $\mathcal{S}_N$ for user modeling. Our paper focuses on leveraging the whole behavior sequence for a comprehensive recommendation. We first illustrate how to explicitly extract short-term and long-term user interests from historical behaviors, and then describe an adaptive way to combine them for joint recommendation. Finally, we introduce a novel dynamic memory network to preserve user's long-term interests for efficient inference effectively. 

\begin{table}[!t]
  \caption{Notations summary.}
  \label{table:notation}
  \centering
\begin{tabular}{c|l}
\hline
\textbf{Notation} & \textbf{Description}  \\ \hline
$u$   & a user \\
$t$   & an item \\
$\mathbf{x}$   & an interaction record \\
$\mathcal{U}$   & the set of users\\
$\mathcal{V}$   & the set of items\\
$\mathcal{S}_n$   & the $n$-th behavior sequence\\
$K$   & the number of candidate items\\  
$N$   &  the number of sliced sequences\\
$L$   & the number of self-attention layers\\
$m$   & the number of memory slots\\
$D$   & the number of embedding dimension\\
$\widetilde{\mathbf{H}}$   & the short-term interest embedding \\
$\widehat{\mathbf{H}}$   & the long-term interest embedding\\
$\mathbf{M}$   & the memory embedding matrix\\ 
$\mathbf{V}$   & the output user embedding\\
\hline
\end{tabular}
\vspace{-0.1in}
\end{table}

\subsection{Recurrent Attention Network}
This subsection introduces the proposed recurrent attention network for short-term interest modeling. Given an arbitrary behavior sequence $\mathcal{S}_n$ as input, an intuitive way to estimate a user's short-term preferences is only consider her/his behaviors within the sequence. 
However, the first few items in each sequence may lack necessary context for effective modeling, because previous sequences are not considered.

To address this limitation, we introduce the notion of recurrence in RNNs into self-attention network and build a sequence-level recurrent attention network, enabling information flow between adjacent sequences. In particular, we use the hidden state computed for last sequence as additional context for next sequence modeling.
Formally, let $\mathcal{S}_{n-1}$ and $\mathcal{S}_{n}$ be two successive sequences, and $\widetilde{\mathbf{H}}^l_{n-1}\in\mathbb{R}^{T\times D}$ denote the $l$-th layer hidden state produced for sequence $\mathcal{S}_{n-1}$. We calculate the hidden state of sequence $\mathcal{S}_n$ as follows. 
\begin{equation}
\begin{aligned}
&\widetilde{\mathbf{H}}_{n}^l= \text{Atten}_{\text{rec}}^l(\widetilde{\mathbf{Q}}_n^l,\widetilde{\mathbf{K}}_n^l,\widetilde{\mathbf{V}}_n^l)=\text{softmax}(\widetilde{\mathbf{Q}}_n^l(\widetilde{\mathbf{K}}_n^l)^\top)\widetilde{\mathbf{V}}_n^l,\\
&\widetilde{\mathbf{Q}}_n^l = \widetilde{\mathbf{H}}_n^{l-1}\widetilde{\mathbf{W}}_Q^\top, \text{\quad and \quad} \widetilde{\mathbf{K}}_n^l = \mathbf{H}_n^{l-1}\widetilde{\mathbf{W}}_K^\top,\\
&\widetilde{\mathbf{V}}_n^l =  \mathbf{H}_n^{l-1}\widetilde{\mathbf{W}}_V^\top,\\
&\mathbf{H}_n^{l-1}=\widetilde{\mathbf{H}}_n^{l-1}\mathbin\Vert \text{SG}(\widetilde{\mathbf{H}}_{n-1}^{l-1}), 
\end{aligned}
 \label{eq1}
\end{equation}
where $\text{Atten}_{\text{rec}}^l(\cdot, \cdot,\cdot)$ is the $l$-th layer self-attention network, in which the query, key and value matrices are denoted by $\mathbf{Q}$, $\mathbf{K}$ and $\mathbf{V}$, respectively. The input of the first layer is the sequence embedding matrix $\mathbf{X}_n=[\mathbf{x}_{n,1},\ldots,\mathbf{x}_{n,T}]\in\mathbb{R}^{T\times D}$. Intuitively, the attention layer calculates a weighted sum of embeddings, where the attention weight is computed between query $i$ in $\mathcal{S}_n$ and value $j$ obtained from previous sequences. The function $\text{SG}(\cdot)$ stands for stop-gradient from previous hidden state $\widetilde{\mathbf{H}}_{n-1}^{l-1}$, and $\mathbin\Vert$ denotes concatenation. In our case, we use the extended context as key and value and adopt three linear transformations to improve the model flexibility, where  $\{\widetilde{\mathbf{W}}_Q,\widetilde{\mathbf{W}}_K,\widetilde{\mathbf{W}}_V\}\in\mathbb{R}^{D\times D}$ denote the parameters. 
The extended context not only provides precious information for recovering the first few items, but also allows our model to capture the dependency across sequences. 
In practice, instead of computing the hidden states from scratch at each time point, we cache the hidden state of last sequence for reuse. Besides, masking strategy and positional embeddings are also included to avoid the future information leakage problem~\cite{yuan2019simple} and capture sequential dynamics~\cite{kang2018self,wang2019encoding}. The final short-term interest embedding is defined as $\widetilde{\mathbf{H}}_n=\widetilde{\mathbf{H}}_n^{L}$.

\subsection{Long-term Attention Network}
In this subsection, we present another attention network for long-term interest modeling. With the recurrent connection mechanism defined in Eq.~\ref{eq1}, our model can capture correlations between adjacent sequences for interest modeling. 
However, longer-range dependencies beyond successive sequences may still be ignored, since the recurrent connection mechanism is limited in capturing longer-range correlations~\cite{sodhani2018training,sukhbaatar2015end}. Hence, additional architecture is needed to effectively capture long-term user preferences.   

To this end, we maintain an external memory matrix $\mathbf{M}\in\mathbb{R}^{m\times D}$ to explicitly memorize a user's long-term preferences, where $m$ is the number of memory slots. Each user is associated with a memory. Ideally, the memory complements with the short-term interest modeling, with the aim to capture dependencies beyond adjacent sequences. We leave how to effectively update the memory in later section and now focus on how to extract long-term interests from the memory. Specifically, let $\mathbf{M}^l\in\mathbb{R}^{m\times D}$ denote the $l$-th layer memory matrix, we estimate the long-term hidden state of sequence $\mathcal{S}_n$ using another self-attention network as
\begin{equation}
\begin{aligned}
\widehat{\mathbf{H}}_{n}^l&= \text{Atten}^l(\widehat{\mathbf{Q}}_n^l,\widehat{\mathbf{K}}_n^l,\widehat{\mathbf{V}}_n^l),\\
\widehat{\mathbf{Q}}_n^l,\widehat{\mathbf{K}}_n^l,\widehat{\mathbf{V}}_n^l&=\widehat{\mathbf{H}}_n^{l-1}\widehat{\mathbf{W}}_Q^\top,\mathbf{M}^{l-1}\widehat{\mathbf{W}}_K^\top,\mathbf{M}^{l-1}\widehat{\mathbf{W}}_V^\top. \\
\end{aligned}
 \label{eq2}
\end{equation}
Similarly, $\text{Atten}^l(\cdot,\cdot,\cdot)$ is a self-attention network. It takes the last layer hidden state $\widehat{\mathbf{H}}_n^{l-1}$ as query and uses the layer-wise memory matrix $\mathbf{M}^{l-1}$ as key (value). The input of the query is $\mathbf{X}_n$. By doing so, the output hidden state $\widehat{\mathbf{H}}^l_n\in\mathbb{R}^{T\times D}$ is a selective aggregation of $m$ memory blocks, where the selection weight is query-based and varies across different queries. 
$\{\widehat{\mathbf{W}}_{Q},\widehat{\mathbf{W}}_{K},\widehat{\mathbf{W}}_{V}\}$ are trainable transformation matrices to improve model capacity. 
Since the memory is maintained to cache long-term user interests that beyond adjacent sequences, 
we refer to the above attention network as long-term interest modeling.
The final long-term interest embedding for sequence $\mathcal{S}_k$ is denoted as $\widehat{\mathbf{H}}_n=\widehat{\mathbf{H}}_n^{L}$.

\subsection{Neural Gating Network}
After obtaining the short-term and long-term interest embeddings, the next aim is to combine them for comprehensive modeling.
Considering that a user's future intention can be influenced by early behaviors, while short-term and long-term interests may contribute differently for next-item prediction over time~\cite{ma2019memory}, we apply a neural gating network to adaptively control the importance of the two interest embeddings. 
\begin{equation}
\begin{aligned}
\mathbf{V}_{n}&= \mathbf{G}_n\odot \widetilde{\mathbf{H}}_n + (1 - \mathbf{G}_n)\odot \widehat{\mathbf{H}}_{n},\\
\mathbf{G}_n&=\sigma(\widetilde{\mathbf{H}}_n\mathbf{W}_{short} + \widehat{\mathbf{H}}_n\mathbf{W}_{long}),
\end{aligned}
 \label{eq3}
\end{equation}
where $\mathbf{G}_n\in\mathbb{R}^{T\times D}$ is the gate matrix learned by a non-linear transformation based on short-term and long-term embeddings. $\sigma(\cdot)$ indicates the sigmoid activation function, $\odot$ denotes element-wise multiplication, and $\mathbf{W}_{short}$, $\mathbf{W}_{long}\in\mathbb{R}^{D\times D}$ are model parameters. The final user embedding $\mathbf{V}_n\in\mathbb{R}^{T\times D}$ is obtained by a feature-level weighted sum of two types of interest embeddings controlled by the gate. 

\subsection{Dynamic Memory Network}
\label{memory:continual}
In this subsection, we describe how to effectively update the memory $\mathbf{M}$ to preserve long-term user preferences beyond adjacent sequences.
One feasible solution is to maintain a fixed-size FIFO memory to cache long-term message. 
This strategy is sub-optimal for user modeling due to two reasons.
First, the oldest memories will be discarded if the memory is full, whether it is important or not. This setting is reasonable in NLP task~\cite{rae2019compressive} as two words that too far away in the sentence are often not correlated. But it is not held in recommendation because behavior sequence is not strictly ordered~\cite{yuan2019simple} and users often exhibit monthly or seasonal periodic behaviors. Second, the memory is redundant and not effectively utilized, since user interests in practice is often bounded in tens~\cite{li2019multi}.   

To avoid these limitations, we propose to abstract a user's long-term interests from the past actively. 
Assume the model was processed sequence $\mathcal{S}_n$, then the memory is updated as 
\begin{equation}
\begin{aligned}
\mathbf{M}^l \leftarrow f_{abs}^l(\mathbf{M}^l,\widetilde{\mathbf{H}}^{l}_{n-1}),
\end{aligned}
 \label{eq4}
\end{equation}
where $f_{abs}^l: \mathbb{R}^{(m+T)\times D}\rightarrow \mathbb{R}^{m\times D}$ is the $l$-th layer abstraction function. It takes the old memory and the context state $\widetilde{\mathbf{H}}^{l}_{n-1}$ as input and updates memory $\mathbf{M}^l$ to represent user interests. 
In theory, $f_{abs}$ requires to effectively preserve the primary interests in old memories and merges contextual information. 
Basically, $f_{abs}$ can be trained with the next-item prediction task end-to-end. Nevertheless, memories that differ from the target item may be discarded. Therefore, we consider train the abstraction function with an auxiliary attention-based reconstruction loss as follows.
\begin{equation}
\begin{aligned} \mathcal{L}_{ae}&= \min \sum_{l=1}^L ||\text{attent}_{rec}^l(\widetilde{\mathbf{Q}}^l,\widetilde{\mathbf{K}}^l, \widetilde{\mathbf{V}}^l)-\text{attent}_{rec}^l(\widetilde{\mathbf{Q}}^l,\widehat{\mathbf{K}}^l,\widehat{\mathbf{V}}^l)||^2_F\\
\widetilde{\mathbf{Q}}^l&=\widetilde{\mathbf{H}}^l_n, \ \widetilde{\mathbf{K}}^l=\widetilde{\mathbf{V}}^l=\mathbf{M}^l\mathbin\Vert\widetilde{\mathbf{H}}_{n-1}^l, \ \widehat{\mathbf{K}}^l=\widehat{\mathbf{V}}^l=\mathbf{M}^l,
  \label{eq5}
 \end{aligned}
\end{equation}
where $\text{atten}_{rec}^l(\cdot, \cdot,\cdot)$ is the self-attention network defined in Eq.~\eqref{eq1}. We reuse the recurrent attention network but keep the parameters fixed and not trainable here. We employ the hidden state $\widetilde{\mathbf{H}}^l_n$ of $\mathcal{S}_n$ as query for two attention networks. The first attention outputs a new representation for the query via a weighted sum from the old and new memories, while the second from the abstracted memories. 
By minimizing the reconstruction loss, we expect the primary interests can be extracted by $f_{abs}$ as much as possible. Note that we consider a lossy objective here because the information that is no longer attended to in $\mathcal{S}_n$ can be discarded in order to capture the shifting of user interests to some extent. 

\subsubsection{Implementation of abstraction function $f_{abs}$} 
We parameterize $f_{abs}$ with the dynamic routing method in CapsNet~\cite{sabour2017dynamic} for its promising results in capturing user's diverse interests in recommendation~\cite{li2019multi}. Suppose we have two layers of capsules, we refer capsules from the first layer and the second layer as primary capsules and interest capsules, respectively. The goal of dynamic routing is to calculate the values of interest capsules given the primary capsules in an iterative fashion. In each iteration, given primary capsules vectors $\mathbf{x}_i$ (input vector), $i\in\{1,\ldots,T+m\}$ and interest capsules $\bar{\mathbf{x}}_j$ (output vector), $j\in\{1,\ldots,m\}$, the routing logit $b_{ij}$ between primary capsule $i$ and interest capsule $j$ is computed by
\begin{equation}
\begin{aligned}
b_{ij}=\bar{\mathbf{x}}_j^\top \mathbf{W}_{ij}\mathbf{x}_i,
  \label{eq5}
 \end{aligned}
\end{equation}
where $\mathbf{W}_{ij}$ is a transformation matrix. Given routing logits,
$\mathbf{s}_j$ is computed as weighted sum of all primary capsules
\begin{equation}
\begin{aligned}
\mathbf{s}_{j}=\sum_{i=1}^{m+T}\alpha_{ij}\mathbf{W}_{ij}\mathbf{x}_i,
  \label{eq6}
 \end{aligned}
\end{equation}
where $\alpha_{ij}=\exp(b_{ij}) / \sum_{j'=1}^{m+T}\exp ( b_{ij'})$ is the connection weight between primary capsule $i$ and interest capsule $j$.
Finally, a non-linear "squashing" function~\cite{sabour2017dynamic} is proposed to obtain the corresponding vectors of interest capsules as 
\begin{equation}
\begin{aligned}
\bar{\mathbf{x}}_{j}=\text{squash}(\mathbf{s}_j)=\frac{\mathbin\Vert\mathbf{s}_j\mathbin\Vert^2}{1+\mathbin\Vert\mathbf{s}_j\mathbin\Vert^2}\frac{\mathbf{s}_j}{\mathbin\Vert\mathbf{s}_j\mathbin\Vert}.
  \label{eq7}
 \end{aligned}
\end{equation}
The routing process between Eq.~\eqref{eq5} and Eq.~\eqref{eq7} usually repeats three times to converge. When routing finishes, the output interest capsules of user $u$ are then used as the memory, i.e., $\mathbf{M}=[\bar{\mathbf{x}}_1,\ldots,\bar{\mathbf{x}}_m]$. 

\subsection{Model Optimization}
As the data is derived from the user implicit feedback, we formulate the learning problem as a binary classification task. Given the training sample $(u,t)$ in a sequence $\mathbf{S}_n$ with the user embedding vector $\mathbf{V}_{n,t}$ and target item embedding $\mathbf{x}_t$, we aim to minimize the following negative likelihood 
\begin{equation}
\begin{aligned}
\mathcal{L}_{like} & = -\sum_{u\in\mathcal{U}}\sum_{t\in\mathcal{S}_n}\log P(\mathbf{x}_{n,t}|\mathbf{x}_{n,1},\mathbf{x}_{n,2},\cdots,\mathbf{x}_{n,t-1})  \\
& = -\sum_{u\in\mathcal{U}}\sum_{t\in\mathcal{S}_n} \log \frac{\exp(\mathbf{x}_t^\top\mathbf{V}_{n,t})}{\sum_{j\in\mathcal{V}}\exp(\mathbf{x}_{j}^\top\mathbf{V}_{n,t}))}.
  \label{eq9}
 \end{aligned}
\end{equation}
The loss above is usually intractable in practice because the sum operation of the denominator is computationally prohibitive. Therefore, we adopt a negative sampling strategy to approximate the softmax function in experiments. When the data volume is large, we leverage Sampled Softmax technique~\cite{covington2016deep,jean2014using} to further accelerate the training. Note that Eqs.~\eqref{eq9} and~\eqref{eq5} are separately updated in order to preserve long-term interests better. Specifically, we first update Eq.~\eqref{eq9} by feeding a new sequence and then updating the abstraction function's parameters by minimizing Eq.~\eqref{eq5}.

%% file: 4.experiment.tex
\begin{table}[!t]
  \caption{The dataset statistics. 
  }
  \begin{tabular}{l|cccc}
    \hline
    Dataset &\#Users &\#Items  & $T$ &$K$\\
     \hline
     MovieLens &6,040 &3,952 &20 &10\\
    Taobao & 987,994 &  4,162,024 &20 &10\\
    JD.com &1,608,707  &378,457 &20 &10\\
    XLong &20,000  &3,269,017 &50 &20\\
  \hline
\end{tabular}
\label{table1}
\vspace{-0.1in}
\end{table}

\begin{table*}
\small
\centering
\caption{Sequential recommendation performance over three benchmarks. $*$ indicates the model only use the latest behavior sequence for training; otherwise, the whole behavior sequence. The second best results are underlined.}
\begin{tabular}{lccccccccc}
    \hline
     \multirow{2}*{\textbf{Models}} &\multicolumn{3}{c}{\textbf{MovieLens}} &\multicolumn{3}{c}{\textbf{Taobao}} &\multicolumn{3}{c}{\textbf{JD.com}}\\
    &HR@10 &HR@50 &NDCG@100 &HR@50 &HR@100 &NDCG@100 &HR@10 &HR@50 &NDCG@100\\
    \hline
    \textbf{GRU4Rec}$^*$ &17.69 &43.13 &16.90 &10.42 &14.01 &4.23 &27.65 &38.73 &23.40\\
    \textbf{Caser}$^*$ &18.98 &45.64 &17.62 &13.71 & 16.51 &6.89 &29.27 &40.16 &24.25\\
    \textbf{SASRec}$^*$ &21.02 &47.28 &19.05 &16.41 &22.83 &9.23 &33.98 &44.89 &27.41\\
    \hline
    \textbf{GRU4Rec} &19.78 &47.40 &18.75 &13.48 &16.53 &5.81 &35.28 &47.52 &27.64\\
    \textbf{Caser} &20.80 &48.12 &19.28 &15.55 &17.91 &7.35 &36.76 &49.13 &28.35\\
    \textbf{SASRec} &22.96 &50.09 &20.36 &20.47 &24.48 &9.84 &38.99 &52.64 &31.32\\
    \hline
    \textbf{SHAN} &21.34 &49.52 &19.55 &18.87 &21.94 &8.73 &37.72 &50.55 &29.80\\
    \textbf{HPMN} &22.84 &50.54 &19.77 &19.98 &24.37 &9.66 &39.14 &53.22 &32.24\\
    \textbf{SDM} &\underline{23.42} &\underline{51.26} &\underline{20.44} &\underline{21.66} &\underline{25.42} &\underline{10.22} &\underline{40.68} &\underline{55.30} &\underline{34.82}\\
    \textbf{DMAN} &\textbf{25.18} &\textbf{53.24} &\textbf{22.03} &\textbf{24.92} &\textbf{29.37} &\textbf{11.13} &\textbf{44.58} &\textbf{58.82} &\textbf{36.93}\\
    \hline
    \textbf{Improv.} &7.51\% &3.86\% &7.77\% &15.05\% &15.53\% &8.90\% &9.58\% &6.36\% &6.05\%\\
    \hline
\end{tabular}
\label{table2}
\end{table*}

\subsection{Datasets}
We conduct experiments over four public benchmarks. Statistics of them are summarized in Table~\ref{table1}. MovieLens~\footnote{https://grouplens.org/datasets/movielens/1m/} collects users' rating scores for movies. JD.com~\cite{lv2019sdm} is a collection of user browsing logs over e-commerce products collected from JD.com. Taobao~\cite{zhu2018learning} and XLong~\cite{ren2019lifelong} are datasets of user behaviors from the commercial platform of Taobao. The behavior sequence in XLong is significantly longer than other three datasets, thus making it difficult to model.

\subsection{Baselines}
To evaluate the performance of DMAN, we include three groups of baseline methods. First, traditional sequential methods. To evaluate the effectiveness of our model in dealing with long behavior sequence, three state-of-the-art recommendation algorithms for sequences with a normal length have been employed, including {\bf GRU4Rec}~\cite{tang2018personalized}, {\bf Caser}~\cite{kang2018self} and {\bf SASRec}~\cite{he2017neural}. Second, long sequential methods. To evaluate the effectiveness of our model in extracting long-term user interests with dynamic memory, we include {\bf SDM}~\cite{lv2019sdm} and {\bf SHAN}~\cite{ying2018sequential}, which are tailored for modeling long behavior sequences. To evaluate the effectiveness of our model in explicitly capturing user's short-term and long-term interests, we also set {\bf HPMN}~\cite{ren2019lifelong} as a baseline. It is based on the memory network. Thrid, DMAN variants. To analyze the contribution of each component of DMAN, we consider three variants. DMAN-XL discards the long-term attention network to verify the effectiveness of capturing long-term interests. DMAN-FIFO adopts a FIFO strategy to validate the usefulness of the abstraction function in extracting primary interests. DMAN-NRAN replaces the recurrent attention network with vanilla attention network to demonstrate the effectiveness of extending context for effective user modeling.

\subsection{Experimental Settings}
We obtain the behavior sequence by sorting behaviors in chronological order. Following the traditional way~\cite{kang2018self}, we employ the last and second last interactions for testing and validation, respectively, and the remaining for training. We follow the widely-adopted way~\cite{li2017neural,lv2019sdm} and split the ordered training sequence into $L$ consecutive sequences. The maximum length of a sequence is $T$. The statistics of four datasets are listed in Table~\ref{table1}. We repeatedly run the model five times and report the average results. 

\subsubsection{Evaluation metrics} For each user in the test set, we treat all the items that the user has not interacted with as negative items. To estimate the performance of top-$K$ recommendations, we use Hit Rate (HR$@K$) and Normalized Discounted Cumulative Gain (NDCG$@K$) metrics, which are widely used in the literature~\cite{he2017neural}. 

\subsubsection{Parameter settings} 
For baselines, we use the source code released by the authors, and their hyper-parameters are tuned to be optimal based on the validation set. To enable a fair comparison, all methods are optimized with the number of samples equals $5$ and the number of embedding dimensions $D$ equals $128$. We implement DMAN with Tensorflow and the Adam optimizer is utilized to optimize the model with learning rate equals $0.001$. The batch size is set to $128$ and the maximum epoch is $8$. The number of memory slots $m$ and attention layers $L$ are searched from $\{2,4,6,8,10,20,30\}$ and $\{1,2,3,4,5\}$, respectively.

\begin{table}
\centering
  \caption{Performance on long user behavior data XLong.}
  \begin{tabular}{cccc}
    \hline
     \textbf{Method} &\textbf{Recall@200} &\textbf{Recall@500}\\
    \hline
    \textbf{GRU4Rec}$^*$ &0.079 &0.098 \\
    \textbf{Caser}$^*$ &0.084 &0.105\\
    \textbf{SASRec}$^*$ &0.105 &0.123\\
    \hline
    \textbf{GRU4Rec} &0.046 &0.063 \\
    \textbf{Caser} &0.023 &0.041\\
    \textbf{SASRec} &0.061 &0.096\\
    \hline
    \textbf{SHAN} &0.091 &0.115\\
    \textbf{HPMN} &0.118 &0.136 \\
    \textbf{SDM} &0.107 &0.129\\
    \textbf{DMAN} &\textbf{0.132} &\textbf{0.163}\\
    \hline
\end{tabular}
\label{table3}
\end{table}

\subsection{Comparisons with SOTA}
In this section, we compare our model with different baselines. Tables~\ref{table2} and~\ref{table3} report the results. In general, we have three aspects of observations. 

\subsubsection{Influence of modeling long behavior sequence for traditional sequential methods} From Table~\ref{table2}, we observe that GRU4Rec, Caser, and SASRec improve their performance when considering longer behavior sequence. Therefore, modeling longer behavior sequence has proved to be effective for user modeling. Besides, different sequential modules have varied abilities in handling long behavior sequence. Specifically, SASRec, GRU4Rec, and Caser improve 24.74\%, 29.36\%, and 13.42\% on Taobao in terms of $\text{HR}@50$, while SASRec consistently performs the best. It indicates the ability of self-attention network in extracting sequential patterns, and also validates our motivation to extend self-attention network for long behavior sequence modeling. 

\subsubsection{Comparison with baselines on general datasets} As shown in Table~\ref{table2}, our model DMAN achieves better results than baselines across three datasets. In general, long sequential models perform better than traditional sequential methods, excepting SASRec. SASRec performs better than SHAN and comparable to HPMN in most cases. This further implies the effectiveness of self-attention network in capturing long-range dependencies. The improvement of SDM over SASRec shows that explicitly extract long-term and short-term interests from long sequence is beneficial. Considering DMAN and SDM, DMAN consistently outperforms SDM over all evaluation metrics. This can be attributed to that DMAN utilizes a dynamic memory network to actively extract long-term interests into a small set of memory blocks, which is easier for the attention network to effectively attend relative information than from a long behavior sequence.

\subsubsection{Comparison with baselines on long behavior dataset}
Table~\ref{table3} summarizes the results of all methods on XLong, where the length of behavior sequence is larger than $1000$ on average. Obviously, DMAN significantly outperforms other baselines. Compared with the findings in Table~\ref{table2}, one interest observation is that traditional sequential methods, i.e., GRU4Rec, Caser, and SASRec, performs poorly when directly handling long behavior sequence, and lose to long sequential models in all cases. These results demonstrate the necessity of developing new architectures tailored for long behavior sequence modeling. Another unexpected observation is that HPMN outperforms SDM on average. It further implies the ineffectiveness of attention network in attending relative messages over long sequence. By equipping attention network with the dynamic memory, our model allows us to actively update user's long-term interests in the memory and outperforms HPMN. 

\begin{table}
\centering
  \caption{Ablation study of DMAN.}
  \begin{tabular}{cccc}
    \hline
     \textbf{Dataset} &\textbf{Method} &\textbf{Recall@100} &\textbf{NDCG@100}\\
    \hline
    \multirow{4}*{\textbf{Taobao}}
    &\textbf{DMAN-XL} &0.237 &0.094 \\
    &\textbf{DMAN-FIFO} &0.263 &0.108\\
    &\textbf{DMAN-NRNA} &0.257 &0.104\\
    &\textbf{DMAN} &0.293 &0.111\\
    \hline

    \multirow{4}*{\textbf{XLong}}
    &\textbf{DMAN-XL} &0.021 &0.013 \\
    &\textbf{DMAN-FIFO} &0.036 &0.017\\
    &\textbf{DMAN-NRAN} &0.043 &0.019\\
    &\textbf{DMAN} &0.054 &0.022\\
    \hline
\end{tabular}
\label{table4}
\end{table}

\subsection{Ablation Study} We also conduct experiments to investigate the effectiveness of several core components of the proposed DMAN. Table~\ref{table4} reports the results on two representative datasets. Obviously, DMAN significantly outperforms the other three variants. The substantial difference between DMAN and DMAN-XL shows that recurrent connection is not enough to capture user's long-term interests. The improvement of DMAN over DMAN-FIFO validates that the proposed abstraction function is effective to extract user's primary long-term interests. Besides, DMAN outperforms DMAN-NRAN in general, which verifies the usefulness of extending current context with previous hidden sequence state for short-term interest extraction. 

\begin{figure*}[htbp]
\centering
 
\subfigure[Memory slots $m$]{
    \begin{minipage}[t]{0.33\linewidth}
        \centering
        \includegraphics[width=5cm,height=2.9cm]{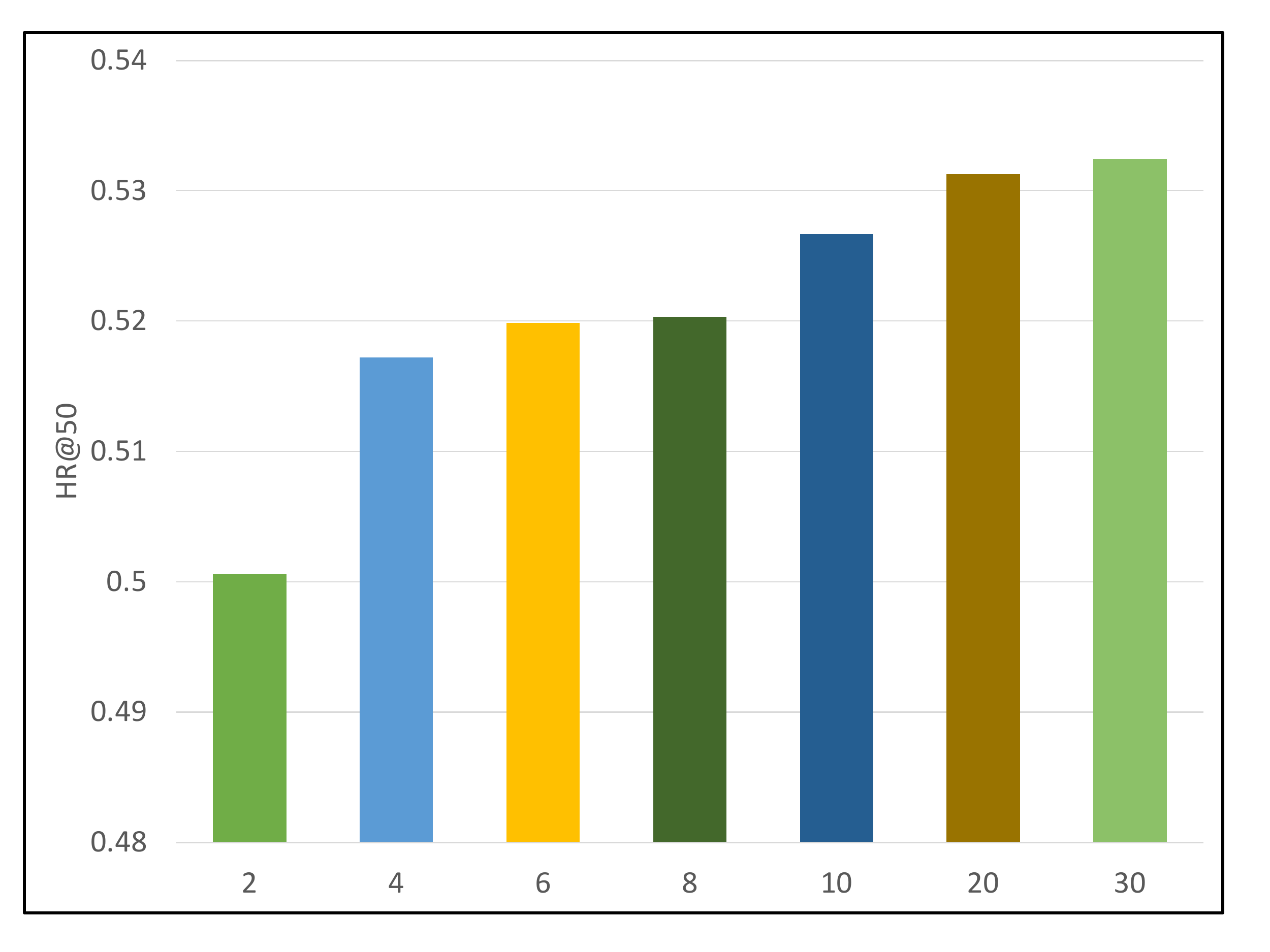}\\
        \vspace{0.02cm}
    \end{minipage}%
}%
\subfigure[Layer size $L$]{
    \begin{minipage}[t]{0.33\linewidth}
        \centering
        \includegraphics[width=6cm,height=2.9cm]{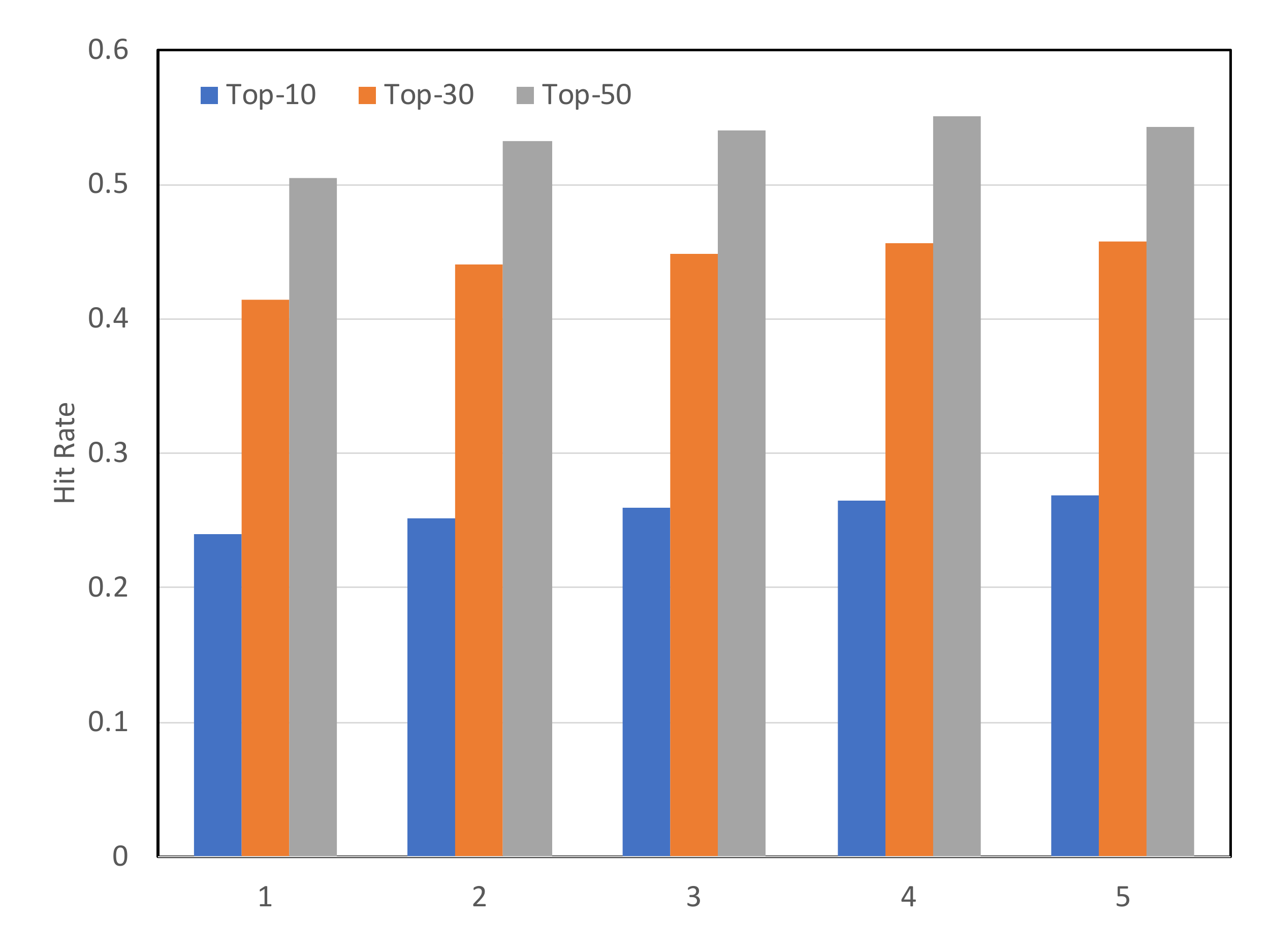}\\
        \vspace{0.02cm}
    \end{minipage}%
}%
\subfigure[Learning curve]{
    \begin{minipage}[t]{0.33\linewidth}
        \centering
        \includegraphics[width=5cm,height=3.1cm]{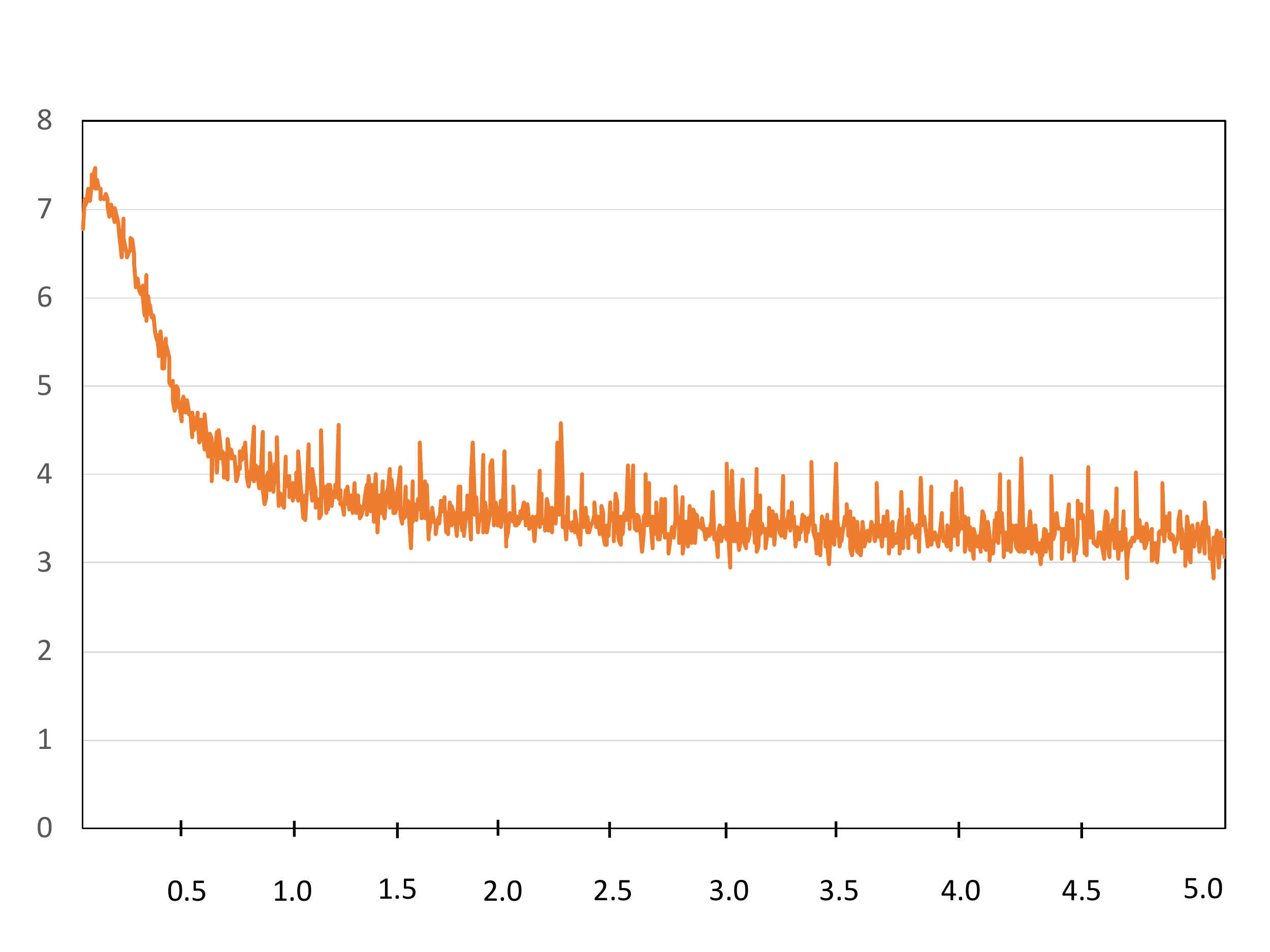}\\
        \vspace{0.02cm}
    \end{minipage}%
}%
\centering
\caption{The proposed DMAN analysis}
\vspace{-0.2cm}
\label{figure2}
\end{figure*}

\subsection{Hyper-parameter Analysis}
We further study the impacts of our model w.r.t. the number of memory slots $m$ and attention layers $L$ on MovieLens. As we can see in Figure~\ref{figure2}(a), DMAN achieves satisfactory results when $m=20$ and the gain slows down with less than 2\% improvement when $m$ further increases. In experiments, we found 20 is enough for MovieLens, Taobao, JD.com and XLong. 
From Figure~\ref{figure2}(b), we observe that the number of attention layers has positive impacts in our model. 
To trade-off between memory costs and performance, we set $L=2$ for all datasets since it already achieves satisfactory results. Besides, we also plot the learning curve of DMAN on Taobao dataset in Figure~\ref{figure2}(c), we can observe that DMAN converges quickly after about 2 epochs. Similar observations have been observed on other datasets. Specifically, DMAN tends to converge after 2 epochs on Taobao, JD.com and XLong datasets, while 50 epochs for MovieLens data. These results demonstrate the training efficiency of our model.

%% file: 2.relatework.tex
\subsection{General Recommendation} 
Early recommendation works largely focused on explicit feedback~\cite{koren2008factorization}. The recent research focus is shifting towards implicit data~\cite{li2017collaborative,hu2008collaborative}. The typical examples include collaborative filtering~\cite{sarwar2001item,schafer2007collaborative}, matrix factorization techniques~\cite{koren2009matrix}, and factorization machines~\cite{rendle2010factorization}. The main challenge lies in representing users or items with latent embedding vectors to estimate their similarity. Due to their ability to learn salient representations, neural network-based models~\cite{guo2017deepfm,su2009survey,tan2019deep} are also attracted much attention recently. Some efforts adopt neural networks to extract side attributes for content-aware recommendation~\cite{kim2016convolutional}, while some aim to equip matrix factorization with non-linear interaction function~\cite{he2017neural} or graph convolutional aggregation~\cite{wang2019neural,liu2019single}. In general, deep learning-based methods perform better than traditional counterparts~\cite{sedhain2015autorec,xue2017deep}.

\subsection{Sequential Recommendation} Sequential recommendation takes as input the chronological behavior sequence for user modeling. Typical examples belong to three categories. The first relies on temporal matrix factorization~\cite{koren2009collaborative} to model user's drifting preferences. The second school uses either first-order ~\cite{rendle2010factorizing,cheng2013you} or hider-order ~\cite{he2016fusing,he2016vista,yan2019cosrec} Markov-chains to capture the user state dynamics. The third stream applies deep neural networks to enhance the capacity of feature extraction~\cite{yuan2019simple,sun2019bert4rec,hidasi2018recurrent}. For example, Caser~\cite{tang2018personalized} applies CNNs to process the item embedding sequence, while GRU4Rec~\cite{hidasi2015session} uses gated recurrent unit GRU for session-based recommendation. 
Moreover, SASRec~\cite{kang2018self} employs self-attention networks~\cite{vaswani2017attention} to selectively aggregate relevant items for user modeling.


However, these methods mainly focus on making recommendations based on relatively recent behaviors. Recently, a few efforts attempt to model long behavior sequence data. For instance, SDM~\cite{lv2019sdm} and SHAN~\cite{ying2018sequential} split the whole behavior sequence into short-term and long-term sequences and then explicitly extract long-term and short-term interest embeddings from them. But they are difficult to capture long-term interests shifting and suffer from high computation complexity.
HPMN~\cite{ren2019lifelong} uses the memory network~\cite{graves2014neural,chen2018sequential} to memorize important historical behaviors for next-item prediction.
Nevertheless, memory network may suffer from long-term dependency forgetting dilemma, as the memory is optimized by recovering the next item. Our model focuses on combing external memory and attention networks for effective long user behavior sequence modeling, which conducts an explicit and adaptive modeling process.


%% file: 5.conclusion.tex
In this paper, we propose a novel dynamic memory-based attention network DMAN for sequential recommendation with long behavior sequence. We truncate a user's overall behavior sequence into a series of sub-sequences and train our model in a dynamic manner. DMAN can explicitly extract a user's short-term and long-term interests based on the recurrent connection mechanism and a set of external memory blocks. To improve the memory fidelity, we derive a dynamic memory network to actively abstract a user's long-term interests into the memory by minimizing a local reconstruction loss. Empirical results on real-world datasets demonstrate the effectiveness of DMAN in modeling long user behavior sequences. 
